\documentclass[twocolumn,prd,aps,showpacs,showkeys,amsmath,amssymb]{revtex4-1}
\usepackage{bm}
\usepackage{color}
\usepackage{amsmath}
\usepackage{amsfonts}
\usepackage{verbatim}
\usepackage{amssymb}
\usepackage{graphicx}
\usepackage{epstopdf}
\usepackage{mathrsfs}
\usepackage{epsfig}
\usepackage{slashed}
\usepackage{bbold}

\providecommand{\U}[1]{\protect\rule{.1in}{.1in}}
\newcommand{\bl}{\boldsymbol}

\newcommand{\ph}{\phantom}

\begin{document}

\title{On the Equivalence of Spacetimes, the Cartan-Karlhede Algorithm}
\author{Thiago M. Mergulh\~{a}o and Carlos Batista}
\email[]{carlosbatistas@df.ufpe.br}
\affiliation{Departamento de F\'{\i}sica, Universidade Federal de Pernambuco,
Recife, Pernambuco  50740-560, Brazil}

\begin{abstract}
It is well known that in general relativity theory two spacetimes whose metrics are related by a coordinate transformation are physically equivalent. However, given two line elements, it is virtually impossible to implement the most general coordinate transformation in order to check the equivalence of the spacetimes. In this paper we present the so-called Cartan-Karlhede algorithm, which provides a finite sequence of steps to decide whether or not two metrics are equivalent. The point of this note is to illustrate the method through several simple examples, so that the reader can learn the fundamentals and details of the algorithm in practice.
\end{abstract}
\keywords{Equivalence of spacetimes; Cartan algorithm; Coordinate transformations}
\maketitle

\section{Introduction}

In the theory of general relativity, the gravitational field is represented by the metric of the spacetime, a rank two symmetric tensor of Lorentzian signature. One of the main features of this theory is that it is totally covariant, in the sense the equations of motion are written in a way that are invariant under the change of coordinates. This is a quite appealing property, since the observers are totally free to choose how to label the points of the spacetime and the physical results should not depend on these arbitrary choices. However, this desirable feature of the theory comes at a price, the difficulty of knowing whether two metrics are equivalent or not. More precisely, given two line elements in two coordinate systems, it is generally very hard to check if they represent the same physical spacetime. For instance, in the search of new exact solutions for Einstein's vacuum equation two people can find two solutions that look completely different, due to the use of different coordinate systems, but they might represent the same physical spacetime. In the same fashion, while integrating Einstein's equation one might end up with several integration constants. However, in general, many of these constants can be eliminated by a coordinate transformation, with just a few of them being physically relevant. These are the problems that are going to be tackled in the present paper.

For instance, consider the following two line elements:
\begin{widetext}
\begin{equation}\label{Sch1}
  ds^2 = -\left(1-\frac{2M}{r} \right)dt^2 + \left(1-\frac{2M}{r} \right)^{-1}dr^2 + r^2 d\theta^2 + r^2 \sin^2\theta d\phi^2 \,,
\end{equation}
\begin{equation}\label{Sch2}
  d\tilde{s}^2 = - \frac{b^2}{ax^2}(a x^2 -c)e^{2\tau}d\tau^2 +  \frac{4a^3 x^4}{a x^2 -c} dx^2 +  \frac{a^2 c^2 x^4}{4 -c^2 y^2} dy^2
  +  \frac{a^2 c^2 x^4 y^2}{4(y^2 + z^2)^2} \left( z^2 dy^2 + y^2 dz^2 - 2 zy \,dydz  \right)  \,.
\end{equation}
\end{widetext}
The first line element is the well-known Schwarzschild solution in spherical coordinates, which represents the gravitational field outside a static and spherically symmetric distribution of mass, with the total mass being $M$. In its turn, the second line element is not known and has three constant parameters, namely $a$, $b$, and $c$. Although they look completely different, it follows that both metrics represent the same gravitational field whenever $c = 2M$. In particular, this implies that the parameters $a$ and $b$ are meaningless and can be eliminated by a coordinate transformation.

In order to check whether the above line elements represent the same gravitational field, the most naive way to proceed is to look for a coordinate transformation that connect both metrics. However, in general this procedure is unbearable, due to the intricate and nonlinear equations involved. A more clever path would be to compare things that do not depend on the coordinate system. For instance, it is well-known that Schwarzschild solution has a vanishing Ricci tensor, since it is a vacuum solution. Therefore, one should check whether the same is true for the line element (\ref{Sch2}). If the second line element is not Ricci-flat then we can immediately conclude that they are not the same solution. However, it turns out that the line element (\ref{Sch2}) is also Ricci-flat, so that the test is inconclusive.

Another clever idea is to compare curvature invariants, which are scalars constructed exclusively from the Riemann tensor, its derivatives, and the metric. This is interesting because scalars are invariant under coordinate transformations. However, since in general the curvature invariants are not constant, they will depend explicitly on the adopted coordinates, which is somehow complicated to compare. For example, let us compare the curvature invariants of the line elements (\ref{Sch1}) and (\ref{Sch2}). The Ricci scalar is easy, since both spaces are Ricci-flat they both vanish. The next simple curvature invariant is the Kretschamnn scalar, given by $\mathcal{R}_2\equiv R^{\mu\nu\alpha\beta}R_{\mu\nu\alpha\beta}$, where $R_{\mu\nu\alpha\beta}$ denotes the components of the Riemann tensor. For the metrics (\ref{Sch1}) and (\ref{Sch2}) we have
\begin{equation*}
   \mathcal{R}_2 =  \frac{48M^2}{r^6}  \;\; \textrm{ and } \;\;
 \tilde{\mathcal{R}}_2 =\frac{12c^2}{a^6 x^{12}} \,,
\end{equation*}
respectively. Note that we cannot tell whether they are the same, since we do not know the possible relations between the coordinates $r$ and $x$ as well as the relations between the constants $M$, $a$, $b$, and $c$.

A way to scape from the awkward position of comparing scalars in different coordinate systems is to try to write curvature invariants in terms of the curvature invariants themselves and then compare the functional relations between them. For instance, defining the cubic curvature invariant
\begin{equation*}
  \mathcal{R}_3\equiv R^{\mu\nu}_{\ph{\mu\nu}\alpha\beta}R^{\alpha\beta}_{\ph{\alpha\beta}\rho\sigma}R^{\rho\sigma}_{\ph{\rho\sigma}\mu\nu}\,,
\end{equation*}
it follows that for the solutions (\ref{Sch1}) and  (\ref{Sch2}) it is given respectively by
\begin{equation*}
 \mathcal{R}_3 =  \frac{96M^3}{r^9}  \;\; \textrm{ and } \;\;
 \tilde{\mathcal{R}}_3 =\frac{12c^3}{a^9 x^{18}} \,.
\end{equation*}
Thus, in both cases we have the same functional relation between the curvature invariants, namely
\begin{equation*}
  \mathcal{R}_3 = \frac{1}{2\sqrt{3}}  (\mathcal{R}_2)^{2/3}  \;\; \textrm{ and } \;\;
    \tilde{\mathcal{R}}_3 = \frac{1}{2\sqrt{3}}  (\tilde{\mathcal{R}}_2)^{2/3} \,.
\end{equation*}
However, this is not enough for asserting that the metrics are the same. Indeed, they are generally not the same, unless we have specifically $c=2M$. Nevertheless, we can move on and compare the functional relations between other scalars, like $\nabla^\sigma R^{\mu\nu\alpha\beta}\nabla_{\sigma}R_{\mu\nu\alpha\beta}$ and so on. If at least one of the functional relations between the curvature invariants does not coincide, then the solutions are not the same, whereas if all the functional relations are the same there is the possibility that the line elements describe the same geometry. Notwithstanding, even in the latter case one cannot guarantee that the solutions are the same. An example of this statement is given by the so-called VSI spacetimes, standing for vanishing scalar invariants spacetimes, which are defined as the Lorentzian metrics for which all curvature invariants vanish identically \cite{Coley1,Coley2}. An example of a VSI spacetime is provided by the line element
\begin{equation}\label{VSI}
  d\hat{s}^2 = H(u,x,y)du^2 + 2dudv + dx^2 + dy^2 \,,
\end{equation}
in which $H(u,x,y)$ is an arbitrary function of the coordinates $x$, $y$, and $u$. All curvature invariants of the latter metric are zero. For instance,
$\hat{R}= \hat{\mathcal{R}}_2=\hat{\mathcal{R}}_3 = 0$ for this metric. Another spacetime with all curvature invariants vanishing is the Minkowski spacetime, which is flat. However, the line element (\ref{VSI}) and the Minkowski metric do not represent the same geometry. Indeed, while Minkowski spacetime has vanishing curvature, the line element (\ref{VSI}) yields a nonzero Riemann tensor. This example proves that it is possible for two metrics to have all the functional relations between their curvature invariants coinciding and still they represent different gravitational fields. In spite of this dramatic example, in several families of spacetimes one can distinguish its members by looking at the curvature invariants, examples of which are given in Refs. \cite{Poly3D,Wylleman:2019hzj}. In addition,  it is interesting that more recently it has been proposed that the event horizons of black holes could be identified by the vanishing of certain curvature invariants \cite{Horizon}.

Thus, in conclusion, the functional relations between the curvature invariants cannot be used to state that two line elements describe the same spacetime geometry but, on the other hand, can be used to rule out the possibility of two metrics representing the same physical solution. There are several other ways to prove the inequivalence of two line elements. For instance, if two metrics admit a different number of independent Killing vector fields or if the algebra of the Killing vector fields is different (thus leading to different isometry groups) they must be different spaces. Likewise, if the two metrics have different algebraic types (according to the Petrov classification, for example) they must be different. However, if we have two metrics to compare and we have computed the functional relations between the curvature, the isometry group and the algebraic types of the curvature tensor, and all of them coincide, even in this case we cannot state that they represent the same gravitational field.

The question that remains to be answered is: is there an algorithmic way to check whether two line elements written in different coordinate systems are the same? The answer is yes and the procedure for comparing the metrics is called Cartan-Karlhede algorithm \cite{Cartan,Karlhede,Brans,Olver,Wylleman:2019hzj,Milson:2012ry}. The goal of this paper is to show by simple examples how this method is used in practice. In particular, many of the calculations are done in three-dimensional spacetimes, in order to make the exposition more clear. This work does not aim in proving the validity of the method but, rather, in being explicit on the meaning of the steps of the algorithm.

The outline of the article is the following. In Sec. \ref{Sec.Algorithm}, we explain the steps of the Cartan-Karlhede algorithm in an abstract way, pointing out some of the subtleties that we should pay attention to while implementing the procedure. Then, in Sec. \ref{Sec.Examples}, we present several examples illustrating the main features of the method. We restrain ourselves to three-dimensional examples, so that the algorithm can be applied  much more neatly and explicitly. Sec. \ref{Appendix4} is then a more advanced section in which we tackle the four-dimensional problem presented in this introduction, namely the comparison of the four-dimensional line elements (\ref{Sch1}) and (\ref{Sch2}). In order to do so, we briefly introduce some useful tools for applying the algorithm. This section can be skipped by the less experienced reader. Finally, in sec. \ref{Sec.Conclusion} presents the discussion of how to circumvent two of the main difficulties that might appear while applying the method.

\section{Cartan-Karlhede Algorithm}\label{Sec.Algorithm}

Suppose that we are interested in comparing two metrics $g_{\mu\nu}(x)$ and  $\tilde{g}_{ab}(\tilde{x})$ of Lorentzian signature living in an open set of an $n$-dimensional manifold described by the coordinate systems $\{x^\mu\}$ and $\{\tilde{x}^\mu\}$. We would like to check whether these two metrics are equivalent or not, in the sense that there exists a coordinate transformation $x^\mu=x^\mu(\tilde{x})$ such that
\begin{equation*}
  \tilde{g}_{\mu\nu}(\tilde{x}) =
  \frac{\partial x^\alpha}{\partial \tilde{x}^\mu} \frac{\partial x^\beta}{\partial \tilde{x}^\nu}\,g_{\alpha\beta}(x(\tilde{x})) \,.
\end{equation*}
The above equation is a nonlinear set of coupled partial differential equations whose solution by direct integration is generally  unfeasible. The intent of the present section is to present an algorithm which after a finite number of steps will tell us whether the coordinate transformation $x^\mu\rightarrow \tilde{x}^\mu$ exists or not. If the answer is positive, we say that the metrics $g_{\mu\nu}$ and  $\tilde{g}_{\mu\nu}$ describe the same geometry; while if the answer is negative we say that these metrics are not equivalent. In what follows we will assume Lorentzian signature for definiteness, although the procedure applies for any signature.

In order to implement this formalism, the first thing we need to do is to choose a constant symmetric matrix  $\eta_{ab}$  that is non-degenerate and has Lorentzian signature as a quadratic form, the labels $a$ and $b$ range from $0$ to $n-1$. For instance, we can choose $\eta_{ab}$ to be the usual Minkowski metric,
$$ \eta_{ab}=\textrm{diag}(-1,1,1,\cdots,1) \,, $$
although any constant symmetric matrix with the correct signature will do the job. Then, introduce a frame of vector fields $\{\mathbf{e}_a = e_a^{\;\mu}\partial_\mu\}$ such that the inner products between them yield exactly the constant matrix $\eta_{ab}$, namely
\begin{equation}\label{eta}
  g_{\mu\nu}\,  e_a^{\;\mu}\, e_b^{\;\nu} = \eta_{ab}\,.
\end{equation}
Note that once chosen $\eta_{ab}$, the choice of frame is not unique, rather we have a freedom of applying Lorentz transformations $\mathbf{\Lambda}\in O(n-1,1)$. By a Lorentz transformation it is meant that the frame is transformed as
\begin{equation}\label{e'}
  \textbf{e}_a\, \mapsto \, \textbf{e}'_a = \Lambda_a^{\;\;b} \, \textbf{e}_b \,,
\end{equation}
with the matrix $\mathbf{\Lambda}$ obeying
$$ \eta_{ab}  \,\Lambda_c^{\;\;a} \,\Lambda_d^{\;\;b} = \eta_{cd}\,.  $$
This set of matrices form a Lie group of dimension $\frac{1}{2}n(n-1)$, the Lorentz group. Here, the Lorentz transformations can be local, i.e., the matrices $\mathbf{\Lambda}$ can vary from point to point, $\mathbf{\Lambda} = \mathbf{\Lambda}(x) $.

Then, we need to compute the curvature tensor and project its components on the frame. For instance, the component $R_{abcd}$ means
\begin{equation*}
  R_{abcd} \equiv R_{\mu\nu\alpha\beta}\,e_a^{\;\mu}\, e_b^{\;\nu}\,e_c^{\;\alpha}\, e_d^{\;\beta}\,.
\end{equation*}
With this at hand, we should take a look at the components $R_{abcd}$ that are nonconstant and compute the number of functionally independent components, which we will denote by $t_0$. For instance, should the components depend just on $x^1$ and $x^2$, but not on $x^3$, $x^4$, $\cdots$, $x^n$, we would have either $t_0=1$ or $t_0=2$, with the former case happening whenever all components of $R_{abcd}$ depend on the same combination of $x^1$ and $x^2$. In general, we always have $0\leq t_0\leq n$. The next step is to compute the so-called isotropy group, which is comprised by the subset of matrices $\mathbf{\Lambda}$ that preserves the form of the components $R_{abcd}$, namely
\begin{equation*}
  R_{abcd} =   R'_{abcd} \,,
\end{equation*}
where $R'_{abcd}$ denotes the components of the curvature in the transformed frame $\{\textbf{e}'_a\}$. We call this subgroup $H_0$.

As the next step, we need to calculate the derivative of the Riemann tensor and project in our frame, namely we need to compute $\nabla_eR_{abcd}$.  Then,
we define $t_1$ as the number of nonconstant functionally independent components of the set $\{R_{abcd},\nabla_eR_{abcd}\}$. For instance, if $t_0=2$ with $R_{abcd}$ depending just on $x^1$ and $x^2$, while $\nabla_eR_{abcd}$ depends on $x^2$ and $x^3$, then we should set $t_1=3$. After this, we must compute the subgroup of $H_0$ that preserves $\nabla_eR_{abcd}$, we denote this subgroup by $H_1$. In other words, $H_1$ is the subgroup of the Lorentz group such that after the frame transformation (\ref{e'}) we have
\begin{equation*}
  R_{abcd} =   R'_{abcd} \; \textrm{ and } \; \nabla_eR_{abcd} = \nabla'_eR'_{abcd} \,.
\end{equation*}
Note that $H_1\subset H_0$.

After this, we do the same procedure for the second derivative of the curvature, $\nabla_e\nabla_f R_{abcd}$. Namely, we compute the number of nonconstant functionally independent components that exists in the set $\{R_{abcd},\nabla_eR_{abcd},\nabla_e\nabla_f R_{abcd}\}$ and call it $t_2$. Then we compute the subgroup of $H_1$ that preserves the form of $\nabla_e\nabla_f R_{abcd}$ when frame transformation (\ref{e'}) is performed. This subgroup is denoted by $H_2$.

One important question is when we should stop the latter procedure, namely at which order of derivative of the curvature tensor. The answer to this question is in the isotropy groups $H_i$ and in the parameters $t_i$. Actually, as defined above, it follows that the numbers $t_i$ are not invariant features of the metric, as they depend on the used frame. Indeed, since we can make arbitrary local Lorentz transformations on the frame these numbers depend on the particular choice of frame. For instance, the component $R_{1212}$ of the curvature tensor might be constant in one frame but through a Lorentz transformation $\mathbf{\Lambda}(x)$ its transformed version $R'_{1212}$ could depend on all the coordinates. Therefore, a much more meaningful number is $\tau_i$, which is defined as the minimum value of $t_i$. More precisely, once computed the curvature and its derivatives we shall look for frames that minimize $t_i$, which is then denoted by $\tau_i$. Metrics that have different values of $\tau_i$ cannot be equivalent, whereas the same cannot be said about $t_i$.  Returning to the question raised at the beginning of this paragraph, we must continue the procedure until we reach an order $q$  such that $\tau_q = \tau_{q-1}$ and $H_{q} = H_{q-1}$. When we reach this point we have a full characterization of the metric $g_{\mu\nu}(x)$ and we can move on to compare with $\tilde{g}_{\mu\nu}(\tilde{x})$. It is worth pointing out that the maximum order $q$ is always finite and cannot exceed $\frac{1}{2}n(n+1)$, as proved by Cartan \cite{Cartan}, although generally $q$ is lower and in some cases we can even anticipate that it must be lower \cite{Karlhede}. In Ref. \cite{Milson:2012ry} it has been explicitly shown that there are 3-dimensional spacetimes such that the algorithm must go up to the maximum theoretical value for $q$. Likewise, in Ref.  \cite{Milson:2007ua} it is proved that the theoretical bound on the value of $q$ is sharp for Petrov type $N$ four-dimensional spacetimes.

One hint of how to check if the minimum value of $t_i$ has been attained is through the use of curvature invariants. For instance, if the Ricci scalar depends just on the coordinate $x^1$ then one can state that $t_0$ is not lower than 1.  Hence, if you manage to find a frame such that $R_{abcd}$ depends just on $x^1$ then you can be sure that $\tau_1=1$.

Once finished the above steps with the metric $g_{\mu\nu}(x)$, let us deal with $\tilde{g}_{\mu\nu}(\tilde{x})$. First, define a frame of vector fields $\{\tilde{\mathbf{e}}_a = \tilde{e}_a^{\;\mu}\tilde{\partial}_\mu\}$ such that
\begin{equation*}
  \tilde{g}_{\mu\nu}\,  \tilde{e}_a^{\;\mu}\, \tilde{e}_b^{\;\nu} = \eta_{ab}\,,
\end{equation*}
where $\eta_{ab}$ is the same matrix used in Eq. (\ref{eta}) when defining the frame $\{\bl{e}_a\}$. Then, compute the components of the curvature in this frame,
\begin{equation*}
  \tilde{R}_{abcd} \equiv \tilde{R}_{\mu\nu\alpha\beta}\,\tilde{e}_a^{\;\mu}\, \tilde{e}_b^{\;\nu}\,\tilde{e}_c^{\;\alpha}\, \tilde{e}_d^{\;\beta}\,.
\end{equation*}
Then we should check whether $R_{abcd} \sim \tilde{R}_{abcd}$ or $R_{abcd}\nsim \tilde{R}_{abcd}$. However, we must be careful with the meaning of the previous equations, since $R_{abcd}$ depend on the coordinates $x^a$, while $\tilde{R}_{abcd}$ depend on $\tilde{x}^a$, so that the comparison is tricky. What is meant by the inequality $\nsim$ is that the functional relations are different. For instance, suppose that
\begin{equation*}
  R_{0101} = x^1 \; \textrm{ and } \;   R_{0202} = 2\, x^1 \,,
\end{equation*}
hence we can write $R_{0101} = 2 R_{0202}$. Thus, if $\tilde{R}_{0101} \neq 2\tilde{R}_{0202}$ we can definitely say that $\tilde{R}_{abcd} \nsim R_{abcd}  $. In order for the metrics to be equivalent it must be possible to find a frame $\{\tilde{\mathbf{e}}_a\}$ such that $\tilde{R}_{abcd} \sim R_{abcd}$, where the symbol $\sim$ denotes that all functional relations agree. If this is not the case for the first adopted frame, we should try to find a Lorentz transformation $\mathbf{\Lambda}\in O(n-1,1)$ such that the equivalence $R_{abcd} \sim \tilde{R}'_{abcd}$ holds, where $\tilde{R}'_{abcd}$ stands for the components of $\tilde{R}_{\mu\nu\alpha\beta}$ in the frame $\{\tilde{\mathbf{e}}'_b\}$ obtained from the initial frame $\{\tilde{\mathbf{e}}_b\}$ through a suitable Lorentz transformation as follows:
\begin{equation}\label{etil'}
   \tilde{ \textbf{e}}'_a = \Lambda_a^{\;\;b} \, \tilde{\textbf{e}}_b \,.
\end{equation}
If we cannot find a frame $\{\tilde{\mathbf{e}}'_b\}$  such that $R_{abcd} \sim \tilde{R}'_{abcd}$, then we conclude that the metrics $g_{\mu\nu}$ and $\tilde{g}_{\mu\nu}$ are not equivalent. Otherwise, if we manage to find a frame  $\{\tilde{\mathbf{e}}'_b\}$ such that $R_{abcd} \sim \tilde{R}'_{abcd}$, we need to compare the first derivative of the curvature tensor.

If $\tilde{\nabla}'_e\tilde{R}'_{abcd} \nsim \nabla_e R_{abcd}$, we need to try to find a frame $\{\tilde{\mathbf{e}}''_b\}$ that is connected to $\{\tilde{\mathbf{e}}'_b\}$ through a Lorentz transformation contained in $H_0$ such that $\tilde{\nabla}''_e\tilde{R}''_{abcd} \sim \nabla_e R_{abcd}$ holds. The requirement that the Lorentz transformation belongs to $H_0$ is to guarantee that $\tilde{R}'_{abcd} = \tilde{R}''_{abcd}$, so that $R_{abcd} \sim \tilde{R}''_{abcd}$  whenever $R_{abcd} \sim \tilde{R}'_{abcd}$ holds. Note that, due to the relation $R_{abcd} \sim \tilde{R}'_{abcd}$, we have $H_0 = \tilde{H}_0$. If we cannot manage to find such a frame, the metrics are not equivalent. On the other hand, if we can find this frame we must continue and compare the second derivative of the curvature. If we manage to equate the curvature and its derivatives of both metrics up to order $q$, where $q$ is the maximum order that we need to go in the procedure with $g_{\mu\nu}$, we conclude that $g_{\mu\nu}$ and $\tilde{g}_{\mu\nu}$ describe the same geometry.

At this point, it is worth stressing that while computing the functional relations one needs to register which components are constant and the values of these constants. More precisely, if one has $R_{0101} = R_{0202}$ and $\tilde{R}_{0101} = \tilde{R}_{0202}$, but $R_{0101}$ is a constant function while $\tilde{R}_{0101}$ is not, it follows that $R_{abcd} \nsim \tilde{R}_{abcd}$. Analogously, if
$R_{0101} = R_{0202}$ and $\tilde{R}_{0101} = \tilde{R}_{0202}$ but $R_{0101} = 1$ while  $\tilde{R}_{0101} = 2$ then we also have $R_{abcd} \nsim \tilde{R}_{abcd}$.

The lists $\{\tilde{\tau}_i,\,\tilde{H}_i\}$ and $\{\tau_i,\, H_i\}$ can be compared as a quick check of wether is there a chance of the metrics being equivalent.
Note, however, that this comparison is not necessary, since these lists are already determined by functional relations of the curvature components. In other words, if we manage to find a frame $\{\tilde{\mathbf{e}}_a\}$ in which the functional relations of the components $\tilde{R}_{abcd}$ and its derivatives coincide with those $R_{abcd}$ and its derivatives, it follows as a consequence that the lists  $\{\tilde{\tau}_i,\,\tilde{H}_i\}$ and $\{\tau_i,\, H_i\}$ also coincide. It seems that this fact have, so far, not been stressed in the literature \cite{Karlhede,McNutt}.

Below we sum up the steps that must be followed in the Cartan-Karlhede algorithm in order to check whether the metrics $g_{\mu\nu}$ and $\tilde{g}_{\mu\nu}$ describe the same geometry.
\begin{enumerate}
  \item Define a constant metric $\eta_{ab}$ and find a frame $\{\mathbf{e}_a\}$ for the metric $g_{\mu\nu}$ whose inner products yield $\eta_{ab}$.
  \item  Compute the curvature components and its derivatives in the previous frame. Also, for each order of derivative, compute $\tau_i$ and $H_i$ the number of functionally independent components that appear up to order $i$ and the isotropy group up to order $i$.
  \item The maximum derivative order that we need to go is the $q^{th}$ order such that $\tau_q = \tau_{q-1}$ and $H_{q}=H_{q-1}$.
  \item Use the Lorentz group degrees of freedom to try to find a frame $\{\tilde{\mathbf{e}}_a\}$ for the metric $\tilde{g}_{\mu\nu}$ such that the functional relations between the components of the Riemann tensor $\tilde{R}_{abcd}$, and its derivatives, coincide with the analogous ones associated to the metric $g_{\mu\nu}$ up to the $q^{th}$ derivative order. If this is not possible the metrics are not equivalent, whereas if one manages to find such a frame we conclude that  there exists a coordinate transformation connecting $g_{\mu\nu}$ and $\tilde{g}_{\mu\nu}$.
\end{enumerate}
In particular, note that, by definition, we must have the following relations:
\begin{align*}
& 0\leq \tau_0 \leq \tau_1 \leq \cdots  \tau_q \leq n \,, \\
&  H_q \subset H_{q-1} \subset \cdots  H_1 \subset H_0 \subset O(n-1,1)\,,
\end{align*}
where $n$ is the dimension of the spacetime. It is worth stressing that although the algorithm tells us whether there exists a coordinate transformation connecting the line elements, it does not provide the actual coordinate transformation.

Note that since the functional relations between the curvature components up to order $q$ in the derivative fully characterize the geometry, one must be able to obtain all geometric information from those. For instance, the isometry group could be extracted. In particular, its dimension is given by
\begin{equation}\label{Isometry}
  D_{\textrm{Isometry}} = (n-\tau_q) + \dim(H_q)\,.
\end{equation}
Recall that if a metric does not depend on some coordinate $z$, namely if $z$ is a cyclic coordinate, then the spacetime is invariant under the translation $z\rightarrow z+ z_0$, where $z_0$ is some constant. We then say that there exists a symmetry on the spacetime, and this symmetry is generated by the vector field $\partial_z$, which is then called a Killing vector field. The isometry group is the set of transformations comprised of all symmetries of the spacetime. Technically, the isometry group is a Lie group whose Lie algebra is formed by all Killing vector fields. Since $\tau_q$ is the number of ``relevant'' coordinates in the geometry, it follows that there are other $(n-\tau_q)$ coordinates that are cyclic. Thus, the dimension of the isometry group is at least $(n-\tau_q)$. However, this reasoning deals only with the transitive part of the isometry group. Indeed, the isometry group at a point $o$ of the manifold can always be decomposed as the product of the transitive subgroup, whose flow changes the point $o$, and the isotropy subgroup, whose flow does not move the point $o$ \cite{Stephani}. This explains why we need to add the term $\dim(H_q)$ in Eq. (\ref{Isometry}). As an example, consider the isometry group of the flat 2-dimensional Euclidean space, whose line element in cartesian coordinates is $ds^2= dx^2+dy^2$. Since $x$ and $y$ are cyclic coordinates, $\partial_x$ and $\partial_y$ are obvious Killing vectors. The flow of these Killing vectors yield a translation in the coordinates $x$ and $y$ respectively. So, in particular, their flows do not leave the origin $(x=0,y=0)$ static. However, besides these Killing vectors we also have $x\partial_y - y\partial_x$, which generates rotations around the origin and, therefore, leave it still. In other words, $x\partial_y - y\partial_x$ generates the isotropy group at the origin.

Up to now, the procedure has been described in a formal and abstract way. In order to make each of the steps more clear, in the next section we will present several simple examples that show explicitly how to apply the algorithm. However, before doing so, let us point out that, in some cases, instead of starting with a randomly chosen frame $\{\mathbf{e}_a\}$ for $g_{\mu\nu}$, it is better to start with a specific frame related to some geometrical structure of $g_{\mu\nu}$. For instance, suppose that $g_{\mu\nu}$ allows a time-like Killing vector field $\bl{K}$, then we can define the Lorentz frame to be such that $\bl{e}_t \propto \bl{K}$, this is a way to partially fix a canonical frame. In this case, if $\tilde{g}_{\mu\nu}$ does not have a time-like Killing vector field the metrics cannot be equivalent. On the other hand, if $\tilde{g}_{\mu\nu}$ has a time-like Killing vector $\tilde{\bl{K}}$, we should choose  $\tilde{\bl{e}}_t \propto \tilde{\bl{K}}$.  This is a coordinate-independent way of trying to fix a canonical frame. The advantage of doing so is that the isotropy freedom that is generally necessary in order to match the functional relations is decreased. These canonical frames can also be fixed by the so-called principal null directions of the Weyl tensor, which are associated to the Petrov classification. More on the latter way of proceeding will be left to Sec. \ref{Appendix4}. Note, however, that these partially fixed frames might not be the ones that minimize $t_i$, in which case they would not be suitable frames for computing $\tau_i$.


\section{Applying the Algorithm in some Examples}\label{Sec.Examples}

For the sake of illustrating the use of the algorithm, in this section we will apply it to several examples. However, instead of considering the dimension four, which is the most physically relevant, we shall tackle three-dimensional examples. This choice is to decrease the number of curvature components that could obfuscate the essentials of the algorithm. This is advantageous because in three dimensions ($n=3$) the Weyl tensor is identically zero, so that the degrees of freedom of the Riemann tensor are totally contained in the Ricci tensor $R_{\mu\nu}$. Thus, the curvature can be represented by a symmetric $3\times 3$ matrix. Also, in three dimensions the Lorentz group is 3-dimensional, while in four dimensions ($n=4$) the Lorentz group has dimension six, so that things are much easier when $n=3$.


First, let us compare the 3-dimensional line element
\begin{equation}\label{3D0}
  ds^2 = -(1+ \alpha\,x)\,dt^2 + dx^2 + dy^2 \,,
\end{equation}
with the following line elements:
\begin{align}
   d\tilde{s}^2 &= -(1+ \alpha\,\tilde{x})\,d\tilde{t}^{\,2}  + \tilde{y}\,d\tilde{x}^2 + d\tilde{y}^2 \,, \label{3D1} \\
   d\breve{s}^2 &= -e^{2\beta \breve{x}}\,d\breve{t}^{\,2} + d\breve{x}^2 + \breve{y}^{2\gamma} d\breve{y}^2 \,,\label{3D4} \\
 d\hat{s}^2 &= -(1 -  \hat{x}^2)\,d\hat{t}^{\,2}  + d\hat{x}^2 + (1 - \hat{x}^2)\,d\hat{y}^2 \,,\label{3D2} \\
 d\bar{s}^2 &= -\left( 1+ \bar{t} + e^{\bar{x}}\right) \,d\bar{t}^{\,2}  + \beta e^{2\bar{x}}d\bar{x}^2  \label{3D3}\\
 & \quad\quad  + 2\beta  e^{\bar{x}}\,d\bar{t} \, d\bar{x} + 4 \bar{y}^2\, d\bar{y}^2 \,.  \nonumber
\end{align}
In these expressions $\alpha$, $\beta$, and $\gamma$ are nonzero constants. In order to perform the comparison, we will choose the frame metric to be the 3-dimensional Minkowski metric
$$ \eta_{ab} = \textrm{diag}(-1,1,1) \,. $$
The orthogonal group associated to this metric is the Lorentz group $O(2,1)$, generated by the composition of the following three types of transformations:
\begin{widetext}
\begin{equation*}
  \textsf{B}_1(\sigma_1) = \left[
                     \begin{array}{ccc}
                       \cosh\sigma_1 & \sinh\sigma_1 & 0 \\
                       \sinh\sigma_1 & \cosh\sigma_1 & 0 \\
                       0 & 0 & 1 \\
                     \end{array}
                   \right]\;,\;\;
   \textsf{B}_2(\sigma_2) = \left[
                     \begin{array}{ccc}
                       \cosh\sigma_2 & 0 & \sinh\sigma_2 \\
                       0 & 1 & 0 \\
                       \sinh\sigma_2 & 0 & \cosh\sigma_2 \\
                     \end{array}
                   \right]\;,\;\;
     \textsf{R}(\theta) = \left[
                     \begin{array}{ccc}
                       1 & 0 & 0 \\
                       0 & \cos\theta & -\sin\theta \\
                       0 & \sin\theta & \cos\theta \\
                     \end{array}
                   \right]  \,,
\end{equation*}
where $\sigma_1$,  $\sigma_2$, and $\theta$ are arbitrary real parameters, with $\theta\in[0,2\pi)$. $\textsf{B}_1$ and $\textsf{B}_2$ are boosts in the space-like directions $\bl{e}_1$ and $\bl{e}_2$ respectively, while $\textsf{R}$ is a rotation in the plane $\bl{e}_1\wedge\bl{e}_2$. These transformations generate the part of $O(2,1)$ that is connected to the identity, namely they can be continuously deformed to the identity. In particular, they all have unit determinant. This connected subgroup is denoted by $SO(2,1)$. Besides, we also have the discrete inversion transformations:
\begin{equation*}
 \textsf{T} = \left[
                     \begin{array}{ccc}
                       -1 & 0 & 0 \\
                       0 & 1 & 0 \\
                       0 & 0 & 1 \\
                     \end{array}
                   \right]\;,\;\;
   \textsf{P}_1 = \left[
                     \begin{array}{ccc}
                       1 & 0 & 0 \\
                       0 & -1 & 0 \\
                       0 & 0 & 1 \\
                     \end{array}
                   \right]\;,\;\;
    \textsf{P}_2 = \left[
                    \begin{array}{ccc}
                       1 & 0 & 0 \\
                       0 & 1 & 0 \\
                       0 & 0 & -1 \\
                     \end{array}
                   \right] \,.
\end{equation*}
The elements of  $O(2,1)$ are given by the composition of these six matrices. More precisely, the most general element of $O(2,1)$ can be written in one of the following forms:
\begin{equation}\label{LorentzBraches}
  \textsf{B}_1 \,\textsf{B}_2 \, \textsf{R} \;,\;\;\;  \textsf{T} \,\textsf{B}_1\,\textsf{B}_2\, \textsf{R} \;,\;\;
  \textsf{P}_1 \,\textsf{B}_1\,\textsf{B}_2\, \textsf{R} \;,\;\;\;
  \textsf{P}_2\,\textsf{B}_1\,\textsf{B}_2\, \textsf{R} \;,\;\;\;  \textsf{T}\, \textsf{P}_1 \,\textsf{B}_1\,\textsf{B}_2\, \textsf{R} \;,\;\;\;
  \textsf{T}\, \textsf{P}_2 \,\textsf{B}_1\,\textsf{B}_2\, \textsf{R} \;,
\end{equation}
where the dependence on $\sigma_1$,  $\sigma_2$, and $\theta$ have been omitted. Note that we have not considered terms of the form $\textsf{P}_1\textsf{P}_2$, since these are already contained in $\textsf{R}(\theta)$ when $\theta=\pi$. As a side note, notice that although $\textsf{T}\textsf{P}_1$ and $\textsf{TP}_2$ have unit  determinant, they are not connected to the identity.
\end{widetext}

Now, following the procedure described in the previous section, let us characterize the metric $g_{\mu\nu}$ given in Eq. (\ref{3D0}). First, we need to define a Lorentz frame. An obvious one is given by
\begin{equation*}
  \bl{e}_0 = \frac{1}{\sqrt{1+\alpha x}} \partial_t \;,\;\;\;
  \bl{e}_1 =   \partial_x \;,\;\;\;
  \bl{e}_2 =  \partial_y \,.
\end{equation*}
The components of the Ricci tensor in this frame are
\begin{equation*}
  R_{ab} = R_{\mu\nu}e_a^{\;\;\mu}  e_b^{\;\;\nu}  = \frac{\alpha^2}{4(1+\alpha x)^2}\left(
             \begin{array}{ccc}
               -1 & 0 & 0 \\
               0 & 1 & 0 \\
               0 & 0 & 0 \\
             \end{array}
           \right)\,.
\end{equation*}
Note that these components depend just on the coordinate $x$, so that we have $t_0=1$. Note also that $R = \frac{\alpha^2}{2(1+\alpha x)^2}$, so that it is impossible to find a frame such that $t_0=0$. Indeed, should we find a frame such that $t_0=0$ the components $R_{ab}$ would be constant so that the Ricci scalar would also be constant, which is not the case. Thus, $\tau_0=1$. In particular, only two components are nonzero, namely $R_{00}$ and $R_{11}$, and these obey the following functional relation:
\begin{equation}\label{FunRelR0-0}
  R_{00} = - R_{11} \,.
\end{equation}
Now we should find the Lorentz transformations that preserve the components $R_{ab}$, namely the ones such that
\begin{equation*}
  R_{ab} = \Lambda_a^{\;\;c} \,  \Lambda_b^{\;\;d} \, R_{cd}\,.
\end{equation*}
Testing each one of the six branches of Eq. (\ref{LorentzBraches}), for arbitrary parameters $\sigma_1$,  $\sigma_2$, and $\theta$, we find that these transformations are given by
\begin{align}
  &\textsf{B}_1(\sigma_1)  \,,\;\;  \textsf{B}_1(\sigma_1)  \textsf{R}(\pi)   \,,\;\;    \textsf{T} \textsf{B}_1(\sigma_1)  \,, \nonumber \\
&\textsf{P}_1 \textsf{B}_1(\sigma_1) \,,\;\;  \textsf{P}_2 \textsf{B}_1(\sigma_1) \,,\;\;   \textsf{T}  \textsf{P}_1 \textsf{B}_1(\sigma_1)   \,,  \label{H03D}\\
    & \textsf{T}  \textsf{P}_2 \textsf{B}_1(\sigma_1)  \,,\;\; \textsf{T} \textsf{B}_1(\sigma_1) \textsf{R}(\pi)\,, \nonumber
\end{align}
for an arbitrary parameter $\sigma_1$. This is a one-dimensional subgroup with eight disconnected branches, this is the isotropy group $H_0$. Formally, this group can be understood as $H_0 = \mathbb{R}\times Z_2 \times  Z_2 \times Z_2 $, where $\mathbb{R}$ is the group formed by the real numbers with the addition operation, and $Z_2$ is the cyclic group of order 2. Denoting the identity transformation by $I$, the three components of $Z_2$ are formed by $\{I,\textsf{T}\}$, $\{I,\textsf{P}_1\}$, and $\{I,\textsf{P}_2\}$, where it is worth recalling that $\textsf{P}_1\textsf{P}_2 = \textsf{R}(\pi)$.

Now, let us evaluate the components of the derivative of the curvature tensor in our frame,
more precisely, $\nabla_a R_{bc}$. Computing them, one verifies that the only nonvanishing components are the following two:
\begin{equation*}
  \nabla_1 R_{00} = -\nabla_1 R_{11} = \frac{\alpha^3}{2(1+ \alpha \,x)^3}\,.
\end{equation*}
Moreover, since these components also depend just on the coordinate $x$, we have $t_1=1$ and $\tau_1=1$. Regarding the indices $bc$ in the object $\nabla_a R_{bc}$, the tensor has the same structure of $R_{bc}$, which would lead to the same isotropy group. However, the existence of nonzero components with the index $a=1$ implies that a change of sign in $\bl{e}_1$, namely the parity transformation $\textsf{P}_1$ and the rotation $\textsf{R}(\pi)$, will change the components   $\nabla_a R_{bc}$ by a sign and, more importantly, $\textsf{B}_1(\sigma_1)$ will not preserve the form of $\nabla_a R_{bc}$ for arbitrary $\sigma_1$, but rather just for the value $\sigma_1=0$, which is the identity transformation. Summing up, the isotropy group up to this order is comprised of the following discrete set of elements:
\begin{equation*}
 I\,,\;\; \textsf{T}  \,,\;\; \textsf{P}_2   \,,\;\; \textsf{T P}_2 \,,
\end{equation*}
where $I$ stands for the identity transformation. Thus, $H_1 = Z_2\times Z_2$, with the cyclic groups given by $\{I,\textsf{T}\}$ and $\{I,\textsf{P}_2\}$. In particular, we have $\dim(H_1)=0$. Note that $\tau_1 =  \tau_0$, but since $H_1\neq H_0$, we have to go to the next order of derivative in the curvature.

Computing the components  $\nabla_a \nabla_b R_{cd}$, we can check that the only nonvanishing components are:
\begin{align*}
  \nabla_0  \nabla_0 R_{11} &= -   \nabla_0  \nabla_0 R_{00} =  \frac{\alpha^4}{4(1+ \alpha \,x)^4} \\
   \nabla_1  \nabla_1 R_{11} &=   - \nabla_1  \nabla_1 R_{00} = \frac{3\alpha^4}{2(1+ \alpha \,x)^4}
\end{align*}
Again, since all the components of $\{R_{ab},\nabla_a  R_{bc},\nabla_a \nabla_b R_{cd}\} $ depend just on the same coordinate $x$, we have $t_2=\tau_2=1$. Moreover, the isotropy group $H_2$ is the set of Lorentz transformations that preserve $\{R_{ab},\nabla_a  R_{bc},\nabla_a \nabla_b R_{cd}\} $ or, equivalently, is the subgroup of $H_1$ that preserves $\nabla_a \nabla_b R_{cd}$. Since all the transformations of $H_1$ also preserve $\nabla_a \nabla_b R_{cd}$, it follows that $H_2 = H_1 = Z_2\times Z_2 $. Once $\tau_2 = \tau_1$ and  $H_2 = H_1$, we can terminate the algorithm at this order.
Thus, the line element (\ref{3D0}) is fully characterized by
the following functional relations between the components of $\{R_{ab},\nabla_a  R_{bc},\nabla_a \nabla_b R_{cd}\} $:
\begin{align}
  R_{00} &= - R_{11} \,,\nonumber \\
  \nabla_1 R_{00} &= -\nabla_1 R_{11} = 4 (R_{11})^{3/2}  \,, \nonumber \\
  \nabla_0  \nabla_0 R_{11} &= -   \nabla_0  \nabla_0 R_{00} =  4 (R_{11})^2\,, \label{Funtional3D-0}\\
   \nabla_1  \nabla_1 R_{11} &= -\nabla_1  \nabla_1 R_{00} =  24 (R_{11})^2\,, \nonumber
\end{align}
with all the components not appearing in the above equation being zero and $R_{11}$ being non-constant. This is the canonical form of the curvature that we need to compare with when testing the equivalence of other line elements.  As a first result, note that the parameter $\alpha$ has no geometrical relevance, since it does not appear at all in the above functional relations. Thus, line elements of Eq. (\ref{3D0}) with different values of $\alpha$ (assuming $\alpha\neq 0$) are all equivalent. Indeed, defining the coordinates
\begin{equation*}
  \texttt{t} = \sqrt{\alpha} \,t \;\;\textrm{ and }\;  \texttt{x} = x -1 + \frac{1}{\alpha} \,,
\end{equation*}
it follows that the line element (\ref{3D0}) becomes
\begin{equation*}
  ds^2 = -(1+ \texttt{x})\,d\texttt{t}^2 +  d\texttt{x}^2 + dy^2 \,,
\end{equation*}
which is just the metric (\ref{3D0}) for $\alpha=1$. Thus, we have proved that all line elements (\ref{3D0}) with nonvanishing $\alpha$ are equivalent to the one with $\alpha=1$.

Now, let us compare the line element (\ref{3D0}) with (\ref{3D1}), (\ref{3D4}), (\ref{3D2}), and (\ref{3D3}). We shall start comparing $g_{\mu\nu}$ with the metric $\tilde{g}_{\mu\nu}$ defined in Eq. (\ref{3D1}). First, note that
\begin{equation*}
  \tilde{\bl{e}}_0 = \frac{1}{\sqrt{1+\alpha \tilde{x}} } \partial_{\tilde{t}} \;,\;\;\;
  \tilde{\bl{e}}_1 =   \frac{1}{\sqrt{\tilde{y}}}  \partial_{\tilde{x}} \;,\;\;\;
  \tilde{\bl{e}}_2 =  \partial_{\tilde{y}}
\end{equation*}
is a frame with the desired inner products.
Computing the components of the Ricci tensor in this frame we find
\begin{equation*}
 \tilde{R}_{ab} = \left(
             \begin{array}{ccc}
               \tilde{R}_{00} & 0 & 0 \\
               0 & \tilde{R}_{22} - \tilde{R}_{00} & (\tilde{R}_{00}\tilde{R}_{22})^{1/2} \\
               0 & (\tilde{R}_{00}\tilde{R}_{22})^{1/2} & \tilde{R}_{22} \\
             \end{array}
           \right)\,,
\end{equation*}
where
\begin{equation*}
  \tilde{R}_{00} = \frac{-\alpha^2}{4\tilde{y} (1+\alpha \tilde{x})^2} \;\; \textrm{ and } \;\;   \tilde{R}_{22} = \frac{1}{4\tilde{y}^2}\,.
\end{equation*}
Thus, we see that $\tilde{R}_{00}$ is functionally independent of $\tilde{R}_{22}$, so that we have $\tilde{t}_0 =2$. However, the frame invariant number is $\tilde{\tau}_0$, which can be different from $\tilde{t}_0$. Computing the Ricci scalar $\tilde{R}$ and the invariant $\tilde{R}^{ab}\tilde{R}_{ab}$ one can check that
$$ \tilde{R}^{ab}\,\tilde{R}_{ab} = \frac{1}{2}\,\tilde{R}^2 \,, $$
which indicate that these curvature invariants are functionally dependent. This could be compatible with $\tilde{\tau}_0=1$. Nevertheless, one can check that $\tilde{R}^{a}_{\;\;b} \tilde{R}^{b}_{\;\;c} \tilde{R}^{c}_{\;\;a}$ is functionally independent from $\tilde{R}$, so that $\tilde{\tau}_0$ must be at least 2. Since we managed to find a frame such that $\tilde{t}_0=2$ we conclude that $\tilde{\tau}_0=2$.Since $\tau_0=1$, we can guarantee right off the bat that $g_{\mu\nu}$ and $\tilde{g}_{\mu\nu}$ describe different geometries. We do not need to bother about using Lorentz transformations in order to match the functional relations, since the parameters $\tilde{\tau}_i$ are invariant under Lorentz transformations. Thus, we will never be able find a frame in which $\tilde{R}_{ab} \sim R_{ab}$.

Then, let us compare $g_{\mu\nu}$ with $\breve{g}_{\mu\nu}$, given in Eq. (\ref{3D4}). Defining
\begin{equation*}
  \breve{\bl{e}}_0 = e^{-\beta \breve{x}} \partial_{\breve{t}} \;,\;\;\;
  \breve{\bl{e}}_1 =     \partial_{\breve{x}} \;,\;\;\;
  \breve{\bl{e}}_2 =  \breve{y}^{-\gamma}\,\partial_{\breve{y}} \,,
\end{equation*}
it follows that $\{ \breve{\bl{e}}_a\}$ is a Lorentz frame associated to $\breve{g}_{\mu\nu}$. Computing the components of the Ricci tensor in such a frame, we find:
\begin{equation*}
 \breve{R}_{ab} =  \beta^2 \left(
             \begin{array}{ccc}
               -1 & 0 & 0 \\
               0 & 1 &  0 \\
               0 & 0 & 0 \\
             \end{array}
           \right)\,.
\end{equation*}
In particular, the components $\breve{R}_{ab} $ obey the relation (\ref{FunRelR0-0}) just as the components $R_{ab}$. In spite of this, we can already conclude that $g_{\mu\nu}$ and $\breve{g}_{\mu\nu}$ are not equivalent, because whereas $\breve{R}_{11}=\beta^2 $ is constant, so that $\breve{t}_0=\breve{\tau}_0=0$, we have found that $R_{11}$ is nonconstant, so that we should write $\breve{R}_{ab} \nsim R_{ab}$. Moreover, note that the metrics $\breve{g}_{\mu\nu}$ for different values of $\beta$ describe different geometries, since no Lorentz transformation can be used to change the set of components $\breve{R}_{ab}$ to a form that does not depend on $\beta$. For instance, it is impossible to find a Lorentz transformation $\{\breve{\bl{e}}_a\}\rightarrow \{\breve{\bl{e}}'_a\}$ such that
\begin{equation*}
\breve{R}'_{ab} =   \left(
\begin{array}{ccc}
-1 & 0 & 0 \\
0 & 1 &  0 \\
0 & 0 & 0 \\
\end{array}
\right)\,.
\end{equation*}

Moving on to the metric $\hat{g}_{\mu\nu}$, defined in Eq. (\ref{3D2}); a Lorentz frame is provided by
\begin{equation*}
  \hat{\bl{e}}_0 = \frac{1}{\sqrt{1-  \hat{x}^2} } \partial_{\hat{t}} \;,\;\;\;
  \hat{\bl{e}}_1 =     \partial_{\hat{x}} \;,\;\;\;
  \hat{\bl{e}}_2 = \frac{1}{\sqrt{1-  \hat{x}^2} } \partial_{\hat{y}} \,.
\end{equation*}
Computing the components of the Ricci tensor in such a frame, we find:
\begin{equation*}
 \hat{R}_{ab} = \left(
             \begin{array}{ccc}
               -\hat{R}_{22} & 0 & 0 \\
               0 & 2(\hat{R}_{22})^2 &0 \\
               0 & 0 & \hat{R}_{22}\\
             \end{array}
           \right)\,,
\end{equation*}
where $\hat{R}_{22} = (1-x^2)^{-1}$. The next step is to try to use the most general Lorentz transformation in order to make $\hat{R}_{ab}$ acquire the functional form of $R_{ab}$, namely $R_{ab} = \textrm{diag}(-R_{11},R_{11},0)$. In order to do so, we must test each of the six branches of the Lorentz group shown in Eq. (\ref{LorentzBraches}). For instance, focusing in the first branch we have
\begin{equation*}
  \hat{\bl{e}}'_a = [\textsf{B}_1(\sigma_1) \textsf{B}_2(\sigma_2) \textsf{R}(\theta)]_a^{\;\;b}\, \hat{\bl{e}}_b \,,
\end{equation*}
whereas the transformed components of the Ricci tensor are
\begin{equation*}
 \hat{R}'_{ab} = [\textsf{B}_1  \textsf{B}_2  \textsf{R} ]_a^{\;\;c} [\textsf{B}_1  \textsf{B}_2  \textsf{R} ]_b^{\;\;d}\,\hat{R}_{cd}\,.
\end{equation*}
The expression for $\hat{R}'_{ab}$ is quite messy, so that here it is shown just the important ones for the argument. For instance,
\begin{equation*}
  \hat{R}'_{22} = \hat{R}_{22} \cosh^2\sigma_2 \,(\cos^2\theta + 2 \hat{R}_{22} \sin^2\theta) \,.
\end{equation*}
In order to achieve $\hat{R}'_{ab} \sim R_{ab}$, one must impose $\hat{R}'_{22}=0$, since $R_{22}=0$. The only solution for this imposition is
\begin{equation}\label{tetahat}
  \theta = \arctan\left( \frac{-1}{2\hat{R}_{22}}\right) \,.
\end{equation}
Then, assuming (\ref{tetahat}) to hold, it follows that
\begin{equation*}
  \hat{R}'_{02} = \sqrt{2} (-\hat{R}_{22})^{2/3} \sinh\sigma_1\cosh\sigma_2 \,.
\end{equation*}
Since one also needs to impose $\hat{R}'_{02} = 0$, one concludes that $\sigma_1=0$. Then, assuming the latter value for $\sigma_1$, it follows that $\hat{R}'_{00}=0$, which is incompatible with the desired equivalence $\hat{R}'_{ab} \sim R_{ab}$. Hence, the first branch of the Lorentz group shown in Eq. (\ref{LorentzBraches}) does not allow a solution for  $\hat{R}'_{ab} \sim R_{ab}$. Likewise, the other five branches yield no solution. Thus, we conclude that $g_{\mu\nu}$ and $\hat{g}_{\mu\nu}$ describe different geometries.

Finally, let us compare the line elements  (\ref{3D0}) and (\ref{3D3}).
First, we need to introduce a Lorentz frame for $\bar{g}_{\mu\nu}$. One option is given by
\begin{align*}
  \bar{\bl{e}}_0 &= \frac{1}{\sqrt{\bar{z}} } \partial_{\bar{t}} \;, \\
  \bar{\bl{e}}_1 &=    \frac{\sqrt{\beta}}{\sqrt{\bar{z} (\bar{z} + \beta) } } \partial_{\bar{t}} + \frac{e^{-\bar{x}} \sqrt{\bar{z}} }{ \sqrt{\beta}\sqrt{\bar{z} + \beta}}  \partial_{\bar{x}} \;, \\
  \bar{\bl{e}}_2 &= \frac{1}{ 2\bar{y}} \partial_{\bar{y}} \,,
\end{align*}
where, for simplification, we have defined $\bar{z}\equiv  \bar{t} + e^{\bar{x}} + 1 $. Computing the components of the Ricci tensor in this frame, we have that the only nonvanishing components are
\begin{equation*}
  \bar{R}_{11} = - \bar{R}_{00} = \frac{1}{4\beta (\bar{z} + \beta)^2}\,.
\end{equation*}
This functional relation is in perfect accordance with the zero order functional relations in Eq. (\ref{Funtional3D-0}). Thus, in order to check whether the line elements describe the same geometry we need to go to higher order. In particular, since $R_{ab} \sim \overline{R}_{ab}$,  it follows that the isotropy group $\bar{H}_0$ is equal to $H_0 = \mathbb{R}\times Z_2 \times  Z_2 \times Z_2 $, i.e. $\bar{H}_0$ is generated by the transformations (\ref{H03D}). Now, let us compute $\bar{\nabla}_{a}\bar{R}_{bc}$. Doing so, we obtain the following nonzero components:
\begin{align*}
  \bar{\nabla}_{0}\bar{R}_{11} &=   - \bar{\nabla}_{0}\bar{R}_{00}  = \frac{1}{2\beta\sqrt{\bar{z}} (\bar{z} + \beta)^3} \,,\\
  \bar{\nabla}_{1}\bar{R}_{00} &=   - \bar{\nabla}_{1}\bar{R}_{11}  =
  \frac{1}{2\beta^{3/2}\sqrt{\bar{z}} (\bar{z} + \beta)^{5/2}}\,.
\end{align*}
These functional relations differ from the ones of $\nabla_{a}R_{bc}$. In particular, here we have $\bar{\nabla}_{0}\bar{R}_{11}$ and $\bar{\nabla}_{0}\bar{R}_{00}$ both different from zero, whereas for the metric $g_{\mu\nu}$ these components vanish, see Eq. (\ref{Funtional3D-0}). However, recall that at order zero we have a nontrivial isotropy group of dimension one. Thus, the frame $\{ \bar{\bl{e}}_a\}$ is not yet fixed by the order zero canonical form for $R_{ab}$. For instance, we can try to use a Lorentz boost $\textsf{B}_1$ in order to try to match $\bar{\nabla}_{a}\bar{R}_{bc}$ and $\nabla_{a}R_{bc}$. Thus, defining
\begin{equation*}
   \bar{\bl{e}}'_a = [\textsf{B}_1(\sigma_1) ]_a^{\;\;b}\, \bar{\bl{e}}_b\,,
\end{equation*}
we find that $\bar{\nabla}'_{0}\bar{R}'_{00}$ and $\bar{\nabla}'_{0}\bar{R}'_{11}$ can be made zero, in accordance with (\ref{Funtional3D-0}), as long as we choose $\sigma_1$ to be
\begin{equation*}
  \sigma_1 = \textrm{arctanh}\left( \frac{\sqrt{\beta}}{\sqrt{\bar{z} + \beta}}\right) \,.
\end{equation*}
Assuming this value for $\sigma_1$ we find that the only nonvanishing components of $\bar{\nabla}'_{a}\bar{R}'_{bc}$ are
\begin{equation*}
  \bar{\nabla}'_{1}\bar{R}'_{00} =   - \bar{\nabla}'_{1}\bar{R}'_{11}  =
  \frac{1}{2\beta^{3/2} (\bar{z}+\beta)^3} = 4 (\bar{R}'_{11})^{3/2}\,,
\end{equation*}
which is in accordance with the functional relations of Eq. (\ref{Funtional3D-0}). So far we have managed to equate the functional relations up to first order derivative of the Riemann tensor, let us finally check the second order. Using the new frame $\{\bar{\bl{e}}'_a\}$ we eventually find that the only nonvanishing components are
\begin{align*}
\bar{\nabla}'_{0} \bar{\nabla}'_{0} \bar{R}'_{11} &=
 - \bar{\nabla}'_{0} \bar{\nabla}'_{0} \bar{R}'_{00}   =
 \frac{1}{4\beta^2 (\bar{z}+\beta)^4} = 4 (\bar{R}'_{11})^2 \,,\\
 \bar{\nabla}'_{1} \bar{\nabla}'_{1} \bar{R}'_{11} &=
 - \bar{\nabla}'_{1} \bar{\nabla}'_{1} \bar{R}'_{00}   =
 \frac{1}{4\beta^2 (\bar{z}+\beta)^4} = 24 (\bar{R}'_{11})^2 \,,
\end{align*}
which agrees perfectly with the functional relations in Eq. (\ref{Funtional3D-0}). Since for the metric $g_{\mu\nu}$ we just need to go up to second order derivative in the curvature, we conclude that $g_{\mu\nu}$ and $\bar{g}_{\mu\nu}$ represent the same geometry. Indeed, it can be checked that performing the coordinate transformation $\{\bar{t},\bar{x},\bar{y}\}\rightarrow \{t,x,y\}$ defined by
$$ \bar{t} = a\, t \,,\;\;  \bar{x} = \log\left( b\,x - a\, t + c  \right)  \,,\;\; \bar{y} = \sqrt{y} \,, $$
where the constants $a$, $b$, and $c$ are given by
$$ a = \alpha^{1/2}\beta^{1/4}\;,\;\; b = \beta^{-1/2} \;,\;\; c = \frac{1-\alpha\beta^{3/2}}{\alpha \beta^{1/2}}  -1 \,, $$
the line element $d\bar{s}^2$, given in Eq. (\ref{3D3}), gets transformed into the line element $ds^2$, given in Eq. (\ref{3D0}). Note that in this example it was necessary to perform a Lorentz transformation in order to compare the components of the curvature and its derivatives. This fact could be expected from the fact that $\partial_t$ is a Killing vector field for the metric $g_{\mu\nu}$ and $\bl{e}_0$ points in the direction of the flow of this symmetry, inasmuch as $\bl{e}_0 \propto \partial_t$. On the other hand, for the metric $\bar{g}_{\mu\nu}$ we have started with a frame such that
$\bar{\bl{e}}_0 \propto \partial_{\bar{t}}$, but now $\bar{t}$ is not a cyclic coordinate, so that $\partial_{\bar{t}}$ is not a Killing vector. Thus, since symmetries are geometrical concepts that are independent of coordinate systems, one could already tell that the frame $\{\bar{\bl{e}}_a\}$ was not the frame $\{\bl{e}_a\}$ written in another coordinate system, so that a Lorentz transformation should be necessary.

As you might have noticed in the previous examples, the tricky part of comparing the metrics stems from the freedom in the choice of the frame. Should we have no such freedom, we would just need to check whether the functional relations between the nonvanishing components of the curvature (and its derivatives) for both metrics are the same or not. Therefore, it would be very handy if we could eliminate the freedom in the choice of the frame. As exemplified before, one way to do so is to look for Killing vectors of the spacetime and use these symmetry directions to form the frame or part of the frame. However, in general, a spacetime has no Killing vector, let alone enough Killing vectors to form a basis, but we should use creativity to seek for geometrical structures that might help on the fixation of a canonical frame. In spite of the latter assertion, this creativity is not mandatory. Indeed, in the 80's and 90's some computer programs have been created to implement the steps of the Cartan-Karlhede algorithm using a computing system called SHEEP, that was created for symbolic tensor calculations. For more details on these programs the reader is refereed to \cite{Aman1,Aman2,McCallum}. As far as the authors could search for, there exists no modern freely available computing package implementing an algorithm for comparing metrics. Thus, the field is open to be explored by the reader. Building such a code would be a really valuable contribution for the community of general relativity. However, a huge price will be paid, in terms of computational time, if the frames are chosen randomly by the computer program. Therefore, in order to make such a program less time-consuming, it would be valuable if the user could fix part of the isotropy freedom by means of geometrical structures. Hence, it would be important for the program to allow the user to use his creativity in order to make the computer task easier. In the next section other important geometrical structures different from Killing vectors are briefly presented, such as the Petrov classification and its associated principal null directions.


\section{A Four-Dimensional Example}\label{Appendix4}

In this section we shall consider the more advanced problem of comparing the four-dimensional line elements (\ref{Sch1}) and (\ref{Sch2}) presented in the introductory section. This example requires more advanced tools of general relativity. Thus, the student of general relativity that is not experienced yet might struggle to follow in detail. Therefore, such a student can skip this section, since the essentials of the method have already been presented in the previous sections. Indeed, the main idea of the present section is to expose the reader to the tools and tricks that are useful in the implementation of the method, so that when a necessity of comparing geometries shows up, the reader has a guide to follow. In this sense, even the beginner student can benefit from acquiring a brief idea about these practical tools. The details can be neglected in a first reading.

In order to compare the line elements (\ref{Sch1}) and (\ref{Sch2}) we first need to define a frame.
An interesting way partially fixing a frame is by means of the so-called principal null directions, which are defined for any spacetime of dimension four whose Weyl tensor is non-vanishing. Given a four-dimensional spacetime one can compute its Weyl tensor $C_{\mu\nu\alpha\beta}$ and try to find the null directions $N^\mu$, with $N^\mu N_\mu = 0$, such that the following constraint holds
\begin{equation}\label{PND}
  N_{[\alpha}C_{\mu]\nu\sigma[\rho}N_{\beta]}N^\nu N^\sigma = 0 \,.
\end{equation}
It turns out that solving this constraint amounts to finding the roots of a fourth order polynomial \cite{chandrasekhar}, so that this algebraic equation admits four solutions, dubbed the four principal null directions (PNDs). These directions can be degenerate or not. The Petrov classification is then an algebraic classification of the Weyl tensor based on the degeneracy of these PNDs \cite{chandrasekhar,art1}. If all four PNDs are distinct we say that the Weyl tensor is type $I$, if two of them coincide and the other two are different we say that it is type $II$, if three coincide we have type $III$, if the four PNDs are the same we have type $N$, while if one pair coincides and the other pair also coincides, but the pairs do not coincide with each other, we say that the Weyl tensor is type $D$. Whenever two metrics have different algebraic types according to the Petrov classification they cannot be equivalent, but if their Petrov type is the same we can use the PNDs to form a frame composed of null vectors.

As mentioned before, we can use any type of frame to start the Cartan-Karlhede algorithm. When using the PNDs to establish the frame, instead of using Lorentz frames, whose inner products give the Minkowski metric, it is useful to adopt a null frame $\{\bl{\ell}, \bl{n}, \bl{m}, \bl{\bar{m}}\}$, in which all vectors are null and the only nonvanishing inner products are
\begin{equation*}
  \ell^\mu n_\mu =-1 \;,\;\; \textrm{and} \;\; m^\mu \bar{m}_\mu =1\,.
\end{equation*}
In other words, the metric components on this frame are given by
\begin{equation}\label{NullFrame}
 \bl{g}(\bl{e}_a,\bl{e}_b) =  \left(
                 \begin{array}{cccc}
                   0 & -1 & 0 & 0 \\
                   -1 & 0 & 0 & 0 \\
                   0 & 0 & 0 & 1 \\
                   0 & 0 & 1 & 0 \\
                 \end{array}
               \right)\,.
\end{equation}
An important point is that in this type of frame it is assumed that $\bl{\ell}$ and $\bl{n}$ are both real, while $\bl{m}$ is complex with its complex conjugate given by  $\bl{\bar{m}}$. Since we are assuming Lorentzian signature, the use of complex vectors is necessary if we insist to work with a frame whose inner products are give by (\ref{NullFrame}). Indeed, the reality conditions of a frame are intimately related to the signature of the metric \cite{Trautman,art1}. In Lorentzian signature, the reality condition of the null frame is given by
$$ \bl{\ell}^\star = \bl{\ell}\;,\;\;\bl{n}^\star = \bl{n}\;, \; \text{and}\; \bl{m}^\star = \bl{\bar{m}}\;,  $$
where the operation $\star$ stands for complex conjugation.

Now, the idea is to fix the frame through the use of PNDs, by choosing the frame vectors  $\bl{\ell}$ and $\bl{n}$ as PNDs. Regarding the Schwarzschild metric, given in Eq. (\ref{Sch1}), it is well-known that its Petrov type is $D$, with the repeated PNDs being
\begin{equation*}
  \bl{N_1} =  f^{-1} \partial_t + f \partial_r \;,\;\;\textrm{and}\;\; \bl{N_2} =  f^{-1} \partial_t - f \partial_r \,,
\end{equation*}
where $f = \sqrt{1- \frac{2M}{r}}$. Taking advantage of these two distinguished null directions, let us choose our null frame to be such that $\bl{\ell}\propto \bl{N_1}$ and  $\bl{n}\propto \bl{N_2}$. More precisely, let us adopt the following frame
\begin{align*}
\bl{\ell} =  \frac{\lambda}{\sqrt{2}}\left( f^{-1} \partial_t + f \partial_r \right) ,\;\,
\bl{n} =  \frac{1}{\lambda \sqrt{2}}\left( f^{-1} \partial_t - f \partial_r \right) ,\\
  \bl{m} =  \frac{e^{i\sigma}}{r\sqrt{2}}\left(  \partial_\theta +i \frac{\partial_\phi}{\sin\theta}   \right) ,\;\,
 \bl{\bar{m}} =  \frac{e^{-i\sigma}}{r\sqrt{2}}\left(  \partial_\theta -i \frac{\partial_\phi}{\sin\theta}  \right).
\end{align*}
Since we have chosen to set $\bl{\ell}\propto \bl{N_1}$ and  $\bl{n}\propto \bl{N_2}$ this fixes partially the frame. In other words, we are using a geometrical structure, namely the PNDs, to define our frame. However, this choice does not fix uniquely the frame, i.e. there is some isotropy remaining. Indeed, if we multiply $\bl{\ell}$ by a real number $\lambda$ and multiply $\bl{n}$ by $\lambda^{-1}$ this will preserve the hypothesis
$\bl{\ell}\propto \bl{N_1}$ and  $\bl{n}\propto \bl{N_2}$, and will also preserve the inner products of the frame. For instance, $\bl{\ell}$ will remain a null vector field, orthogonal to $\bl{m}$ and obeying $\ell^\mu n_\mu = -1$. Likewise, we can multiply $\bl{m}$ by $e^{i\sigma}$ while multiplying $\bl{\bar{m}}$ by $e^{-i\sigma}$ without changing the inner products, preserving the reality condition of the frame and keeping $\bl{\ell}\propto \bl{N_1}$ and  $\bl{n}\propto \bl{N_2}$ valid. Thus, the real parameters $\lambda$ and $\sigma$ in the above expressions for the frame represent the remaining isotropy group. Thus, using the PNDs to define the frame we have reduced the six-dimensional isotropy group associated to a general Lorentz frame to a two-dimensional isotropy group, providing a much easier starting point to apply the Cartan-Karlhede algorithm. In addition to this freedom, we also have the discrete freedoms of interchanging $\bl{\ell}$ by $\bl{n}$ and $\bl{m}$ by $\bl{\bar{m}}$.

Since Schwarzschild spacetime has vanishing Ricci tensor, its curvature reduces to the Weyl tensor. In turn, the only independent nonvanishing component of the Weyl tensor is
\begin{equation}\label{psi2}
  C_{\ell m  n \bar{m}} = \frac{M}{r^3} .
\end{equation}
Although there are other nonvanishing components, like $C_{\ell n\ell n} =-2M/r^3$, these other components must all be proportional to $C_{\ell m \bar{m} n}$ due to the Bianchi identity $C_{[abc]d}=0$ and to the trace-less condition $C^{a}_{\;\;bad}=0$, so that it is unnecessary to consider them in our analysis. Note that this component does not depend on $\lambda$ and $\sigma$, so that we cannot use the isotropy group to fix a value for $C_{\ell m \bar{m} n}$. At this point, we conclude that $t_0 = \tau_0= 1$, due to the dependence on $r$ in the Kretschmann scalar, and $\textrm{dim}(H_0) = 2$, due to the freedom in the choice of $\lambda$ and $\sigma$. Apart from the related components that stem from the identities $C_{[abc]d}=0$, $C^{a}_{\;\;bad}=0$, and $\nabla_{[e}C_{ab]cd}=0$, the only nonzero components of the derivative of the curvature are
\begin{equation}\label{Dpsi2}
  \nabla_\ell  C_{\ell m  n \bar{m}} = -\frac{3M f}{\sqrt{2} \lambda  r^4 } \;,\;\; \nabla_n  C_{\ell m n \bar{m} } = \frac{3M \lambda f}{\sqrt{2}r^4} \,.
\end{equation}
In particular, note that we can use the freedom in the choice of $\lambda$ in order to set $ \nabla_\ell  C_{\ell m  n \bar{m}} = -\nabla_n  C_{\ell m n \bar{m}}$, which amounts to choosing $\lambda=1$. Thus, the dimension of the isotropy is lowered at this order, so that $\textrm{dim}(H_1) = 1$, while $\tau_1=\tau_0=1$. With the choice  $\lambda=1$, we can solve Eq. (\ref{psi2}) for $r$ and substitute this expression for $r$ into (\ref{Dpsi2}), leading to the following functional relation:
\begin{equation*}\label{FunRelSch1}
  \nabla_n  C_{\ell m  n \bar{m}} = \frac{3(C_{\ell m n  \bar{m}})^{4/3}\sqrt{1-2  M^{2/3} (C_{\ell m  n \bar{m}})^{1/3}}}{ \sqrt{2} M^{1/3}}\,.
\end{equation*}
At this point, we already see that line elements with different values of $M$ are not equivalent, since the latter functional relation depends on the parameter $M$ and this dependence cannot be eliminated by means of the isotropy freedom. Moving on to the next order, we see that an example of nonvanishing component is given by
\begin{align*}
  \nabla_\ell \nabla_n  C_{\ell m  n \bar{m} } &= \frac{3M(5M-2r)}{r^6} \\
  &= 15 (C_{\ell m  n \bar{m}})^2 - \frac{6}{M^{2/3}}(C_{\ell m  n \bar{m}})^{5/3}\,,
\end{align*}
just as this component, the other nonzero second order components do not depend on $\sigma$, so that $H_2 = H_1$. In addition, these components depend just on the coordinate $r$, so that $t_2= \tau_2 =1$. Thus, since $\tau_2=\tau_1$ and $H_2=H_1$, we should stop the Cartan-Karlhede characterization at this order. Here we will not list all the nonvanishing second order components, since there are several of them.

Now, let us analyse the line element (\ref{Sch2}). First, we need to set a null frame $\{\tilde{\bl{\ell}}, \tilde{\bl{n}}, \tilde{\bl{m}}, \tilde{\bl{\bar{m}}}\}$ for this line element. In order to perform the comparison with the Schwarzschild geometry, we should choose the null vectors $\tilde{\bl{\ell}}$ and $\tilde{\bl{n}}$ as the principal null directions of the metric (\ref{Sch2}), since this was our choice in the characterization of (\ref{Sch1}). Solving Eq. (\ref{PND}) in this geometry, we find the following principal null directions for the Weyl tensor:
\begin{align*}
 \tilde{\bl{N_1}} &=  \frac{e^{-\tau} }{2b^2}\partial_\tau + \frac{ax^2 - c }{4b a^2 x^3}\partial_x \;,\\
  \tilde{\bl{N_2}} &=  \frac{ a x^2 e^{-\tau} }{ ax^2 -c} \partial_\tau - \frac{b}{2ax} \partial_x \,.
\end{align*}
It turns out that both PNDs are degenerate, so that the line element (\ref{Sch2}) is of Petrov type $D$, just as the Schwarzschild solution, so that there is the possibility of these two geometries being equivalent. Now, let us define a null frame such that $\tilde{\bl{\ell}}\propto \tilde{\bl{N_1}}$, and $\tilde{\bl{n}}\propto \tilde{\bl{N_2}}$. For instance, let us adopt
\begin{align*}
 \tilde{\bl{\ell}} &=  \tilde{\lambda} \,\left( \frac{e^{-\tau} }{2b^2}\partial_\tau +   \frac{ax^2 - c }{4b a^2 x^3}\partial_x \right)\;,\\
  \tilde{\bl{n}} &=  \tilde{\lambda}^{-1}  \left( \frac{ a x^2 e^{-\tau} }{ ax^2 -c} \partial_\tau - \frac{b}{2ax} \partial_x \right)\,, \\
  \tilde{\bl{m}} &=
  \frac{ e^{i\tilde{\sigma} } \sqrt{4 - c^2 y^2}  }{\sqrt{2}\,a c x^2} \left[  \partial_y +
  \left(  \frac{ z  }{  y}  - i \frac{2(y^2 + z^2)}{y^2  \sqrt{4 - c^2 y^2} }  \right)\partial_z \right] \,, \\
\end{align*}
with $\tilde{\bl{\bar{m}}}$ being the complex conjugate of  $\tilde{\bl{m}}$. Note that the remaining isotropy have already been made explicit through the arbitrary real parameters $\tilde{\lambda}$ and $\tilde{\sigma}$ appearing in the basis. Since this geometry is also Ricci-flat ($\tilde{R}_{ab}=0$), its curvature reduces to the Weyl tensor. A nonvanishing component of the Weyl tensor is
\begin{equation}\label{Csch2}
  \tilde{C}_{\tilde{\ell} \tilde{m}  \tilde{n} \tilde{\bar{m}}} = \frac{c}{2 a^3 x^6} \,.
\end{equation}
The other nonvanishing components of $\tilde{C}_{abcd}$ are all proportional to the latter one due to the Bianchi identity and the trace-less property of the Weyl tensor. Concerning the first derivative of the of the curvature, the only ``independent'' nonvanishing components are
\begin{equation*}\label{gbfgf}
 \tilde{\nabla}_{\tilde{\ell}}   \tilde{C}_{\tilde{\ell} \tilde{m}  \tilde{n} \tilde{\bar{m}} } =  \frac{-3bc}{2 a^4 x^8 \tilde{\lambda}}
 \;,\;\; \tilde{\nabla}_{\tilde{n}}   \tilde{C}_{\tilde{\ell} \tilde{m}  \tilde{n} \tilde{\bar{m}} }  = \frac{3c \tilde{\lambda} ( c - a x^2)}{4 b a^5 x^{10} } \,.
\end{equation*}
Thus, making the choice $\tilde{\lambda} = \frac{bx\sqrt{2a}}{\sqrt{ax^2-c}}$ we obtain the functional relation $\tilde{\nabla}_{\tilde{n}}   \tilde{C}_{\tilde{\ell} \tilde{m}  \tilde{n} \tilde{\bar{m}} } =
-\tilde{\nabla}_{\tilde{\ell}}   \tilde{C}_{\tilde{\ell} \tilde{m}  \tilde{n} \tilde{\bar{m}} } $, just as in the Schwarzschild spacetime for the frame $\{\bl{e}_a\}$. More precisely, for this choice of $\tilde{\lambda}$ we obtain
\begin{equation*}\label{gbfgf}
 \tilde{\nabla}_{\tilde{n}}   \tilde{C}_{\tilde{\ell} \tilde{m}  \tilde{n} \tilde{\bar{m}} } =
  -\tilde{\nabla}_{\tilde{\ell}}   \tilde{C}_{\tilde{\ell} \tilde{m}  \tilde{n} \tilde{\bar{m}} }  =
  \frac{3c  \sqrt{ a x^2 - c}}{2 \sqrt{2} a^{9/2} x^{9} } \,.
\end{equation*}
In terms of $ \tilde{C}_{\tilde{\ell} \tilde{m}  \tilde{n} \tilde{\bar{m}}}$, the above relation can be written as
$$
\tilde{\nabla}_{\tilde{n}}   \tilde{C}_{\tilde{\ell} \tilde{m}  \tilde{n} \tilde{\bar{m}} } =
    \frac{3 (\tilde{C}_{\tilde{\ell} \tilde{m}  \tilde{n} \tilde{\bar{m}}})^{4/3}
   \sqrt{ 1- (2c^2 \tilde{C}_{\tilde{\ell} \tilde{m}  \tilde{n} \tilde{\bar{m}}})^{1/3}} }{(2 c^2)^{1/3} } \,.
 $$
This functional relation coincides with the analogous one for the Schwarzschild spacetime as long as we accept the identification $c=2M$. Thus, up to first order in the derivative, we conclude that the metrics can be equivalent, it remains to check the second order relation. The nonvanishing components of the second derivative of the Weyl tensor are the same as the analogous ones for the Schwarzschild spacetime with the frame $\{\bl{e}_a\}$. For instance, the component $ \nabla_{\tilde{\ell}} \nabla_{\tilde{n}}\tilde{C}_{\tilde{\ell} \tilde{m}  \tilde{n} \tilde{\bar{m}}}$ is nonvanishing and given by
$$
\nabla_{\tilde{\ell}} \nabla_{\tilde{n}}\tilde{C}_{\tilde{\ell} \tilde{m}  \tilde{n} \tilde{\bar{m}}} =
\frac{ 3c (5c - 4a x^2)  }{ 4 a^6 x^{12} }\,.
$$
Thus, assuming that $c=2M$, we eventually find that
$$
\nabla_{\tilde{\ell}} \nabla_{\tilde{n}}\tilde{C}_{\tilde{\ell} \tilde{m}  \tilde{n} \tilde{\bar{m}}} =
15 (\tilde{C}_{\tilde{\ell} \tilde{m}  \tilde{n} \tilde{\bar{m}}} )^2 - \frac{6}{M^{2/3}}(\tilde{C}_{\tilde{\ell} \tilde{m}  \tilde{n} \tilde{\bar{m}}} )^{5/3} \,,
$$
which agrees perfectly with the analogous functional relation that we have established for the Schwarzschild spacetime. It can be checked that all other second order functional relations also agree. Thus, we conclude that the line elements (\ref{Sch1}) and (\ref{Sch2}) represent the physical geometry as long as we set $c=2M$. Moreover, as a spin off, we conclude that, since the functional relations of the Schwarzschild spacetime depend on the Mass parameter $M$, different values  of $M$ lead to different geometries. This is the reason why $M$ is a physical parameter. On the other hand, since the parameters $a$ and $b$ that appear in the line element (\ref{Sch2}) do not show up in the functional relations of this line element, they have no physical relevance. In other words, the line elements obtained from Eq. (\ref{Sch2}) by choosing different values for $a$ and $b$ all represent the same geometry.

\section{Discussion and Conclusions} \label{Sec.Conclusion}

In this article we have reviewed the  Cartan-Karlhede algorithm, which is a procedure that allows one to check whether two line elements represent the same geometry, in spite of written in different coordinate systems, by following a finite sequence of steps. The review was performed through the use of simple examples in which the steps of the algorithm have been performed analytically and explicitly. We hope that this provides a useful guide for students and researchers to learn the essentials of the method.

A source of difficulty in the algorithm that, has been glossed over in the above examples is that sometimes it can be quite difficult to solve an equality like $R_{1212} = f(x)$ for the coordinate $x$ in order to find the functional relations in terms of the curvature components themselves, without referring to coordinates, as done in Eq. (\ref{Funtional3D-0}). Moreover, it can happen that one should not solve the relation for a single coordinate but rather for a specific combination of coordinates, as happened in the example in which we have dealt with the three-dimensional line element $\bar{g}_{\mu\nu}$, in which a particularly useful coordinate combination was $\bar{z}\equiv 1 + \bar{t} + e^{\bar{x}} $. This part of the algorithm will generally demand from the user creativity and ingenuity to do smart choices of the components in terms of which the other curvature components will be written. For instance, if $\tau_q$ is the number of functionally independent components of the curvature (and its derivatives), then one should choose $\tau_q$ components of the curvature (and its derivatives) to serve as the ``basis'' in terms of which the other components will be written. Different choices of this ``basis'' can enormously simplify or greatly hamper the establishment of the functional relations. In spite of highlight on the use of creativity, these steps can be done systematically by a computer \cite{Aman1,Aman2,McCallum}, but the computation time can become quite big if the computer turn out to choose an unsuitable path.

As we have stressed throughout the text, the most difficult part of the implementation of the method is working out the remaining isotropy freedom at each order of the method. In order to make things easier in Sec. \ref{Appendix4}, in which we have compared the line elements (\ref{Sch1}) and (\ref{Sch2}), we have started with frames that already incorporated the isotropy freedom through the parameters $\lambda$ and $\sigma$. Then, since the components of the curvature did not depend on $\sigma$, we concluded that this part of the Lorentz group belongs to the isotropy group of the metric. On the other hand, since one of the components depended on $\lambda$, we concluded that the dimension of the Lorentz group associated to the parameter $\lambda$ is not part of the isotropy group of the metric. Moreover, in the latter case, we have chosen $\lambda$ in way to establish a convenient functional relation. For instance, if it is possible to make some curvature component vanish by a suitable choice of a free parameter, we can do so in order to fix the frame. The idea of starting the Cartan-Karlhede procedure algorithm with the Lorentz freedom explicit in the frame is generally handy for establishing the isotropy group at each order. However, in general, allowing the Lorentz group freedom to be explicit from the beginning is unfeasible, due to the complicated expressions for the frame that stem from the fact that a general Lorentz transformation has several parameters. We have been able to adopt this procedure in the four-dimensional example because we have already started with a partially fixed frame, since we have chosen two frame vectors to point along principal null directions. This choice reduced the freedom in the frame, remaining just two dimensions of freedom in the Lorentz group. Thus, one of the main strategies to apply the Cartan-Karlhede algorithm efficiently is to use geometric structures of the metric, namely special directions defined in a coordinate independent form, like symmetry directions and principal null directions, to fix completely or partially the frame with which we start the procedure. In this case it is said that the curvature tensor has been put in a canonical form. As an example, suppose that the Ricci tensor has an eigenvector with constant eigenvalue, then we can use this distinguished eigenvector as one of the frame vectors. One can also use conformal Killing vectors, Killing tensors and Killing-Yano tensors to define special directions or planes along which the frame is aligned. It is worth mentioning that notion of principal null directions of the Weyl tensor are also defined in dimensions greater than four \cite{CMPP,Rev. Coley}, so that this concept can be used to put the curvature in a canonical form in higher dimensions \cite{McNutt}. Keep in mind, however, that in order to compute the parameters $\tau_i$, which are important to decide when the Cartan-Karlhede algorithm should stop, we might have to use a frame different from the one fixed by means of the symmetries and geometrical structures.

\vspace{0.5cm}

\begin{acknowledgments}
C. B. would like to thank Conselho Nacional de Desenvolvimento Cient\'{\i}fico e Tecnol\'ogico (CNPq) for the partial financial support through the research productivity fellowship. Likewise,  C. B. thanks Universidade Federal de Pernambuco for the funding through Qualis A project. We both thank Lode Wylleman and Matthew Aadne for the useful comments on the first version of this paper concerning the relevance of the distinction between $t_i$ and $\tau_i$, which we have tried to make clear in the present version.
\end{acknowledgments}



\end{document}